\documentstyle[12pt]{article}
 
\def\beqra{\begin{eqnarray}} \def\eeqra{\end{eqnarray}} 
\def\beqast{\begin{eqnarray*}} \def\eeqast{\end{eqnarray*}} 
\def\beq{\begin{equation}}      \def\eeq{\end{equation}} 
\def\be{\begin{enumerate}}   \def\ee{\end{enumerate}} 
 
\def\eps{\epsilon} 
 
\def\si{\sigma} 
 
\def\al{\alpha}

\def\BM#1{{\boldmath 
\mathchoice{\hbox{$\displaystyle#1$}} 
           {\hbox{$\textstyle#1$}} 
           {\hbox{$\scriptstyle#1$}} 
           {\hbox{$\scriptscriptstyle#1$}}}}

\def\rta{\rightarrow} 
\def\eqv{\equiv} 
 
\def\pa{\partial} 
 
\def\ch{\@startsection{section}{1}{\z@}{-3ex plus-1ex minus-.2ex}%
        {2ex plus.2ex}{\large\sc{\normalsize}}}

\def\cl{{\cal L}}

\def\raisenot{\raise .5mm\hbox{/}} 
\def\nota{\ \hbox{{$a$}\kern-.49em\hbox{/}}} 
\def\notA{\hbox{{$A$}\kern-.54em\hbox{\raisenot}}} 
\def\notb{\ \hbox{{$b$}\kern-.47em\hbox{/}}} 
\def\notB{\ \hbox{{$B$}\kern-.60em\hbox{\raisenot}}} 
\def\notc{\ \hbox{{$c$}\kern-.45em\hbox{/}}} 
\def\notd{\ \hbox{{$d$}\kern-.53em\hbox{/}}} 
\def\notD{\ \hbox{{$D$}\kern-.61em\hbox{\raisenot}}} 
\def\note{\ \hbox{{$e$}\kern-.47em\hbox{/}}} 
\def\notk{\ \hbox{{$k$}\kern-.51em\hbox{/}}} 
\def\notp{\ \hbox{{$p$}\kern-.43em\hbox{/}}} 
\def\notP{\ \hbox{{$P$}\kern-.60em\hbox{\raisenot}}} 
\def\notq{\ \hbox{{$q$}\kern-.47em\hbox{/}}} 
\def\notQ{\ \hbox{{$Q$}\kern-.61em\hbox{/}}} 
\def\notW{\ \hbox{{$W$}\kern-.75em\hbox{\raisenot}}} 
\def\notz{\ \hbox{{$Z$}\kern-.61em\hbox{\raisenot}}} 
 
\def\notpa{\hbox{{$\partial$}\kern-.54em\hbox{\raisenot}}} 
\def\notveps{\ \hbox{{$\varepsilon$}\kern-.40em\hbox{\raisenot}}} 
 
\def\fo{\hbox{{1}\kern-.25em\hbox{l}}}  

\def\and{\,{\rm and}\,}

\def\7#1#2{\mathop{\null#2}\limits^{#1}}        
\def\5#1#2{\mathop{\null#2}\limits_{#1}}        

\def\BM#1{{\boldmath 
\mathchoice{\hbox{$\displaystyle#1$}} 
           {\hbox{$\textstyle#1$}} 
           {\hbox{$\scriptstyle#1$}} 
           {\hbox{$\scriptscriptstyle#1$}}}} 
 
\def\inbar{\vrule height1.5ex width.4pt depth0pt} 
\def\IB{\relax{\rm I\kern-.18em B}} 
\def\IC{\relax\leavevmode\hbox{\,$\inbar\kern-.3em{\rm C}$}} 
\def\ID{\relax{\rm I\kern-.18em D}} 
\def\IE{\relax{\rm I\kern-.18em E}} 
\def\IF{\relax{\rm I\kern-.18em F}} 
\def\IG{\relax\leavevmode\hbox{\,$\inbar\kern-.3em{\rm G}$}} 
\def\IH{\relax{\rm I\kern-.18em H}} 
\def\II{\relax{\rm I\kern-.18em I}} 
\def\IK{\relax{\rm I\kern-.18em K}} 
\def\IL{\relax{\rm I\kern-.18em L}} 
\def\IM{\relax{\rm I\kern-.18em M}} 
\def\IN{\relax{\rm I\kern-.18em N}} 
\def\IO{\relax\leavevmode\hbox{\,$\inbar\kern-.3em{\rm O}$}} 
\def\IP{\relax{\rm I\kern-.18em P}} 
\def\IQ{\relax\leavevmode\hbox{\,$\inbar\kern-.3em{\rm Q}$}} 
\def\IR{\relax{\rm I\kern-.18em R}} 
\def\sed{\hbox{{\sf S}\kern-.4em\hbox{\sf S}}} 
\def\ZZ{\relax{\sf Z\kern-.4em Z}} 
\def\zed{\hbox{{\sf Z}\kern-.4em\hbox{\sf Z}}} 
\def\smIR{\hbox{{\footnotesize\rm I}\kern-.2em\hbox{\footnotesize\rm R}}} 
\def\smIO{\ \hbox{{\footnotesize\rm I}\kern-.4em\hbox{\footnotesize\bf O}}} 
\def\smIQ{\ \hbox{{\footnotesize\rm I}\kern-.5em\hbox{\footnotesize\bf Q}}} 
\def\IGa{\relax{\rm I}\kern-.18em\Gamma} 
\def\IPi{\relax{\rm I}\kern-.18em\Pi} 
\def\IQt{\relax\leavevmode\hbox{$\kern.3em\inbar\kern-.3em\Theta$}} 
\def\IOm{\relax\hbox{$\kern3.48pt\inbar\kern1.8pt\inbar\kern-5.28pt\Omega$}} 
 
\def\ca#1{\relax\ifmmode {\cal#1} \else$\cal#1$\fi}     
\def\Sf#1{\relax\ifmmode\hbox{\sf#1}\else{\sf#1}\fi}    
\def\fibby{\ifcase\@ptsize                      
                \font\tenrm=cmfib8\or           
                \font\elvrm=cmfib8 scaled\magstephalf\or        
                \font\twlrm=cmfib8 scaled\magstep1 \fi}         
\def\TeXey{\ifcase\@ptsize\or\or                
                \font\twlrm=cmr10 scaled\magstep1       
                \font\twlmi=cmmi10 scaled\magstep1      
                \font\twlit=cmti10 scaled\magstep1      
                \font\twlbf=cmbx10 scaled\magstep1\fi}  
 
\def\ch{\@startsection{section}{1}{\z@}{-3ex plus-1ex minus-.2ex}%
        {2ex plus.2ex}{\large\sc}} 
\def\sch{\@startsection{subsection}{2}{\z@}{-1.5ex plus-1ex minus-.2ex}%
        {1pt plus.2ex}{\sc}} 
\def\ssch{\@startsection{subsubsection}{3}{\z@}{-1ex plus-1ex minus-.2ex}%
        {1pt plus.2ex}{\small\sc}} 
\def\seceq{\@addtoreset{equation}{section}
        \def\theequation{\thesection.\arabic{equation}}}        
 
\def\con{\ifmmode \hbox{\bf*} \else{\bf*}\fi}   
\def\scon{\ifmmode \hbox{\footnotesize\rm\bf*} \else{\footnotesize\rm\bf*}\fi}

\def\0#1{\relax\ifmmode\mathaccent"7017{#1}
        \else\accent23#1\relax\fi}              
 
\def\haf{\frac{1}{2}} 
 
\def\place#1#2#3{\vbox to0pt{\kern-\parskip\kern-7pt 
                             \kern-#2truein\hbox{\kern#1truein #3} 
                             \vss}\nointerlineskip} 
\def\figurecaption#1#2{\kern.75truein\vbox{\hsize=5truein\noindent{\bf Figure 
    \figlabel{#1}:} #2}} 
\def\tablecaption#1#2{\kern.75truein\lower12truept\hbox{\vbox{\hsize=5truein 
    \noindent{\bf Table\hskip5truept\tablabel{#1}:} #2}}} 
\def\boxed#1{\lower3pt\hbox{ 
                       \vbox{\hrule\hbox{\vrule 
                         \vbox{\kern2pt\hbox{\kern3pt#1\kern3pt}\kern3pt}\vrule} 
                         \hrule}}} 
\def\figlabel#1{\global\advance\figureno by 1 
\relax\ifproofmode\ifforwardreference 
\immediate\write1{\noexpand\expandafter\noexpand\def 
\noexpand\csname FIGLABEL#1\endcsname{\the\chapno.\the\figureno?}}\fi\fi 
\global\expandafter\edef\csname FIGLABEL#1\endcsname 
{\the\chapno.\the\figureno?}\iffigurechapternumbers\chapfolio.\fi 
\ifproofmode\llap{\hbox{\marginstyle#1 
\kern1.2truein}}\relax\fi\the\figureno} 
 
\def\place#1#2#3{\vbox to0pt{\kern-\parskip\kern-7pt 
                             \kern-#2truein\hbox{\kern#1truein #3} 
                             \vss}\nointerlineskip}

\def\illustration #1 by #2 (#3){\vbox to #2{\hrule width #1 height 0pt depth 0pt 
                                       \vfill\special{illustration #3}}} 
 
\def\scaledillustration #1 by #2 (#3 scaled #4){{\dimen0=#1 \dimen1=#2 
           \divide\dimen0 by 1000 \multiply\dimen0 by #4 
            \divide\dimen1 by 1000 \multiply\dimen1 by #4 
            \illustration \dimen0 by \dimen1 (#3 scaled #4)}}

\mark{{}{}}   
 
\def\ps@headings{\let\@mkboth\markboth 
\def\@oddfoot{}\def\@evenfoot{}
\def\@evenhead{  \sl \leftmark \hfil}
\def\@oddhead{\hbox{}\hfil \sl \rightmark   }
} 

\def\ps@myheadings{\let\@mkboth\@gobbletwo 
\def\@oddhead{\hbox{}\sl\rightmark \hfil \rm\thepage}%
\def\@oddfoot{}\def\@evenhead{\rm \thepage\hfil\sl\leftmark\hbox {}}%
\def\@evenfoot{}\def\chaptermark##1{}\def\sectionmark##1{}%
\def\subsectionmark##1{}}

\catcode`@=12

\begin{document} 
 
\renewcommand{\theequation}{\arabic{equation}} 
\renewcommand{\thesection}{\Roman{section}} 
\renewcommand{\thesubsection}{\Roman{subsection}} 
 
\def\beqra{\begin{eqnarray}} \def\eeqra{\end{eqnarray}} 
\def\beqast{\begin{eqnarray*}} \def\eeqast{\end{eqnarray*}} 
\def\beq{\begin{equation}}      \def\eeq{\end{equation}} 
\def\be{\begin{enumerate}}   \def\ee{\end{enumerate}} 
\def\notu{\ \hbox{{$u$}\kern-.47em\hbox{/}}} 
 
\hfill{DOE-ER-40757-109} 
 
\hfill{OSU-HEP-98-3}

\hfill{UTEXAS-HEP-98-2} 
 
\vspace{12pt} 
\begin{center} 
 \large{\bf COLLIDER PRODUCTION OF SPIN {$\displaystyle{\BM{\frac{3}{2}}}$} 
QUARKS} 
 
\normalsize 
 
\def\spin{\frac{3}{2}}

\vspace{36pt} 
Duane A. Dicus \\ 
 
\vspace{12pt} 
{\it Center for Particle Physics and Department of Physics\\ 
University of Texas, Austin, Texas 78712} 
 
\vspace{24pt} 
 S. Gibbons and S. Nandi \\ 
 
\vspace{12pt} 
{\it Department of Physics   \\ 
Oklahoma State  University,  Stillwater, OK  } 
\end{center} 
 
\vspace{24pt} 
\abstract{We consider the production of spin $\frac{3}{2}$ quarks in 
hadron-hadron and photon-photon colliders.  The cross sections at LHC energy is 
large enough to observe such exotic quarks up to a mass of a few TeV.} 
 
\baselineskip=21pt 
\section{Introduction} 
 
\indent\indent  The recent discovery of the top quark at the Fermilab Tevatron, 
both by the CDF  and D0 Collaboration has filled the last missing 
ingredient in the fermionic sector of the Standard Model (SM).  The 
measured mass 
and the cross section values are in agreement with those expected from the 
Standard 
Model.  With the luminosity and/or energy upgrade of the Tevatron collider and 
with the commissioning of the LHC, we will be able to explore the  missing 
bosonic ingredient of the SM, namely the Higgs boson.  We will also be able to 
explore the presence of heavy 4th generation fermions and also other exotic 
gauge bosons and fermions.  In this work we consider the production of such an 
exotic particle, namely a spin $\frac{3}{2}$ quark.  Just like an ordinary 
quark, 
we assume it to be a color triplet. 
 
It is not outside the realm of possibility that a  spin $\frac{3}{2}$ quark 
could 
exist as a fundamental particle.  We could also have 
spin $\frac{3}{2}$ bound states of ordinary quarks with gluons or 
the Higgs boson.  There are also theoretical models in which spin $\frac{3}{2}$ 
quarks arise as bound states of three heavy quarks for sufficiently strong 
Yukawa couplings \cite{tay}.  
The masses of these bound states are typically 
expected to be a few TeV.  
A heavy spin $\frac{3}{2}$ quark could also exist as the lightest Regge
recurrences of light spin $\frac{1}{2}$ quarks. These also could exist as
Kaluza-Klein modes in string theory if one or more of the compactification
radii is of the order of the weak scale rather than the Planck scale. Such 
weak compactification in the framework of both string theory or field
theory has recently been popular \cite{sure}.
In this work, we restrict ourselves to the collider 
production of point-like spin $\frac{3}{2}$ color triplet quarks,
 in $pp~,p\bar p$ and 
$\gamma\gamma$ colliders.  The production of spin $\frac{3}{2}$ quarks has 
been 
previously considered by Moussallam and Soni for hadronic collisions 
\cite{mous}. 
 Our analytical results for the gluon fusion subprocess is in 
disagreement with theirs. 
 
Our paper is organized as follows.  In section II, we give the Lagrangian and 
the Feynman rules relevent for the production of spin $\frac{3}{2}$ quarks in 
hadron-hadron and photon-photon colliders.  In section III, we give the 
analytic 
formulae for the various subprocess cross sections and also the total cross 
sections for the $pp$ and $p \bar p$ colliders. 
In section IV we give the results for the photon-photon colliders.  Section V 
contains our conclusions. 
 
\section{Feynman Rules for Spin  $\frac{3}{2}$ Particles}

\indent\indent  The Lagrangian and the equations of motion for a free spin 
$\frac{3}{2}$ particle of mass $M$ can be written as \cite{rari, mold} 
\beqra 
&& \cl = \bar\psi_\al \Lambda_{\al\beta}\;\psi_\beta \\[6pt] 
&& \Lambda_{\al\beta}\;\psi_\beta=0 
\eeqra 
where 
\beqra 
\Lambda_{\al\beta} &=& (i\notpa-M) 
g_{\al\beta}+iA(\gamma_\al\pa_\beta+\gamma_\beta\pa_\al) \nonumber \\[6pt] 
&& + \frac{iB}{2}\,\gamma_\al\,\notpa\,\gamma_\beta+ 
CM\;\gamma_\al\gamma_\beta\\[6pt] 
\nonumber
\eeqra
with $B \eqv 3A^2+2A+1$ and $C \eqv 3A^2+3A+1$.
The parameter $A$ is arbitrary except that  $A\neq-\haf$.  The field $\psi_\al$ 
satisfies the subsidiary conditions 
\beqra 
\gamma_\al\;\psi_\al &=& 0 \\ 
\pa_\al\;\psi_\al &=& 0\;. 
\eeqra 
The Lagrangian (1) is invariant under the point transformation \cite{mold, 
nath} 
\beqra 
\psi_\al \rta\psi'_\al &=& \psi_\al+d\gamma_\al\gamma_\lambda\psi^\lambda \\ 
A\rta A' &=& \frac{A-2d}{1+4d} 
\eeqra 
where $d$ is an arbitrary parameter except $d\neq-\frac{1}{4}$.The propagator
$S_{\al\beta}(p)$ is given by 
\beqra 
S_{\al\beta}(p)&=&\frac{1}{\notp-M} \Bigg[ g_{\al\beta}-\frac{1}{3}\; 
\gamma_\al\gamma_\beta -\frac{2}{3M^2}\, p_\al\,p_\beta \nonumber \\ 
&& -\frac{1}{3M}\left(\gamma_\al p_\beta-\gamma_\beta p_\al\right)\Bigg] 
\nonumber \\ 
&& + \Bigg\{ 
\frac{a^2}{6M^2}\notp\gamma_\al\gamma_\beta-\frac{ab}{3M}\; 
\gamma_\al\gamma_\beta \nonumber \\ 
&& + \frac{a}{3M^2}\; \gamma_\al \,p_\beta + \frac{ab}{3M^2}\,\gamma_\beta 
p_\al\Bigg\} 
\eeqra 
where 
$$a=\frac{A+1}{2A+1}\;,\;b\eqv\frac{A}{2A+1}\;.$$ 
Note that the terms depending on the parameter A disappear on the mass 
shell. Pascalutsa has  proposed that the spin $\frac{3}{2}$ field 
$\psi_\al$ can be redefined so that all the $A$ dependent terms can be absorbed 
in the definition of $\psi_\al$ and no explicit $A$ dependence appears in the 
propagator \cite{pasc}.  In our calculation, as an additional check, we 
used the 
general 
$A$ dependent propagator given by (8) since the cross sections or 
any physical quantity is independent of $A$. 
  Moussallam and Soni took $A=-1$.   The interaction Lagrangian for the 
color triplet spin $\frac{3}{2}$ quarks with the gluons or photon is obatined 
by using the minimal substitution in (1). 
\beq 
\cl_I=g\bar\psi_\al 
\left(\frac{B}{2}\;\gamma^\al\gamma^\mu\gamma^\beta + 
Ag^{\al\mu}\gamma^\beta+A\gamma^\al g^{\mu\beta} 
+g^{\beta\al}\gamma^\mu\right)T_a \psi_\beta A_\mu^a 
\eeq 
where $g$ is the gauge coupling constant, $T_a$are the group generators and 
$A^a_\mu$ are the gauge fields. 
 
\section{Calculation of Cross Sections for Hadron Colliders} 
 
\indent\indent  In this section, we calculate the cross sections for the 
processes 
\beq 
\bar pp\rta Q\bar Q+ anything 
\eeq 
and 
 
\beq 
pp\rta Q\bar Q+ anything 
\eeq 
where $Q$ represents the spin $\frac{3}{2}$ quark. 
 
The subprocesses contributing to both reactions (10) and (11) are 
 
\beq 
g+g\rta Q\bar Q 
\eeq 
and 
 
\beq 
q+\bar q\rta Q\bar Q\;. 
\eeq 
The Feynman diagrams contributing to the gluon-gluon $(gg)$ and 
quark-antiquark $(q\bar q)$ subprocess are shown in fig. 1.  The amplitudes 
for the $t,~u$ and s-channels are 
 
\beqra 
M_t &=& g^2\bar u^\rho(p)\left(\gamma^\mu g^{\rho\al}+A\gamma^\al 
g^{\mu\rho}\right) T_a\eps_{\mu a}(k) \nonumber \\ 
&& \Bigg\{ \frac{1}{\notp-\notk-M} \Bigg[ g^{\al\beta}-\frac{1}{3} \,\gamma^\al 
\gamma^\beta -\frac{2}{3M^2}(p-k)^\al(p-k)^\beta \nonumber \\ 
&& -\frac{1}{3M}\left(\gamma^\al(p-k)^\beta-\gamma^\beta(p-k)^\al\right) \Bigg] 
\nonumber\\ 
&& +\frac{a^2}{6M^2}(\notp-\notk)\gamma^\al \gamma^\beta-\frac{ab}{3M}\, 
\gamma^\al\gamma^\beta \nonumber \\ 
&& + \frac{a}{3M^2}\, \gamma^\al(p-k)^\beta + \frac{ab}{3M^2}\,\gamma^\beta 
(p-k)^\al\Bigg\} \nonumber \\ 
&& \left( \gamma^\nu g^{\beta\si}+A \gamma^\beta g^{\si\nu}\right) T_b\eps_{\nu 
b}(k') v^\si(p') 
\eeqra 
\pagebreak 
 
\beqra 
M_u&=& g^2\bar u^\rho(p)\left(\gamma^\nu 
g^{\rho\al}+A\gamma^\al g^{\nu\rho}\right) T_b\eps_{\nu b}(k') \nonumber \\ 
&& \Bigg\{ \bigg[ g^{\al\beta}-\frac{1}{3}\, \gamma^\al\gamma^\beta - 
\frac{2}{3M^2} (k-p')^\al(k-p')^\beta \nonumber \\ 
&& + \frac{1}{3M}\left(\gamma^\al(k-p')^\beta-\gamma^\beta(k-p')^\al\right) 
 \bigg]\frac{1}{\notk-\notp'-M} \nonumber \\ 
&& +\frac{a^2}{6M^2}(\notk-\notp')\gamma^\al\gamma^\beta-\frac{ab}{3M}\, 
\gamma^\al \gamma^\beta  \nonumber \\ 
&&+ \frac{a}{3M^2}\, \gamma^\al(k-p')^\beta+ 
\frac{ab}{3M^2}\,\gamma^\beta(k-p')^\al\Bigg\}\nonumber \\ 
&& ~~~~ (\gamma^\mu g^{\beta\si}+A\gamma^\beta g^{\si\mu})T_a\eps_{\mu a}(k) 
v^\si 
\eeqra 
 
\beqra 
M_s &=& - g^2 if_{abc} \,\bar u^\rho(p)\gamma^\al 
T_a v^\rho(p')\,\frac{1}{\hat s}\,\eps_{\mu b}(k)\eps_{\nu c}(k') \nonumber \\ 
&& \bigg[(2k+k')^\nu g^{\mu\al}-(k+2k')^\mu g^{\nu\al}+(k'-k)^\al 
g^{\mu\nu}\bigg]\;. 
\eeqra 
We point out here that although $M_t$ and $M_u$ depends on the contact 
transformation parameter $A,\,\sum| M_t^2|$ and $\sum|M_u^2|$ and all the cross 
terms are separately independent of $A$.  $M_t,\;M_u$ and $M_s$ satisfy the 
appropriate  gauge invariance conditions.  The amplitude for the quark 
subprocess 
(13) is given by 
\beq 
M_q = -ig^2\,\frac{1}{\hat s}\,\bar u^\al(p) T_a\gamma^\mu v_\al(p')\,
\bar u(p_1)\gamma_\mu v(p_2)\,. 
\eeq 
Using (12)--(16), the total cross section for the gluon-gluon subprocess is 
obtained to be 
\beqra 
\hat\si(gg\rta Q\bar Q) &=& \frac{\pi \al_s^2}{116,640\hat s}\; \Bigg\{ 60 \ln 
\frac{1+\beta}{1-\beta}\; \bigg[66\,y^2+8y \nonumber \\ 
&& + 886+5,184\,\frac{1}{y}+1,296\,\frac{1}{y^2}\bigg] \nonumber \\ 
&&+\beta\bigg[24\,y^4+1,178\,y^3-13,626\,y^2+11,380\,y \nonumber \\ 
&& - 97,200-602,640\;\frac{1}{y}\bigg]\Bigg\} 
\eeqra 
where $\al_s\eqv g^2/4\pi$ and $y\eqv\hat s/M^2$ and 
$\beta\eqv\sqrt{1-4/y }~$. 
Most of the coefficients in (18) are in disagreement with those of Eq. (14) of 
Moussallam and Soni \cite{mous}; only the first and last term in the first 
square 
bracket and the first term in the second square bracket agree.  
A preliminary version of this paper was sent to the authors of 
ref. \cite{mous}. Professor B.~Moussallam has informed us that he has
found an algebraic error in their calculations. After correcting for that
error, their new results agree with our equation (18) above \cite{mous2}. 
For the 
quark-antiquark subprocess, the total cross section is 
\beq 
\hat\si(q\bar q\rta Q\bar Q)=\frac{\pi\al^2_s}{81\hat s}\, \beta 
\left[\frac{8}{3} y^2-\frac{16}{3}\, y-\frac{16}{3} + 96\, \frac{1}{y}\right] . 
\eeq 
This quark antiquark subprocess was not calculated in Ref. 2. 
 
The total cross sections for the processes (10) and (11) are obtained by 
folding 
in the appropriate quark, antiquark and gluon distributions.  We have used the 
distributions produced by the CTEQ Collaboration evaluated at $q^2=M^2$. 
 
In Fig. 2, we give the contributions to the cross section for spin 
$\frac{3}{2}$ quark $(Q)$ pair production at the Tevatron
($p\bar p$, $\sqrt{s}=1.8$ TeV) due 
to the gluon-gluon and  quark-antiquark subprocesses and also the total cross 
section.  As expected, at this energy, the cross section is dominated by the 
quark-antiquark subprocess. 
 The results  for 
the TeV 2000 ($p\bar p$, $\sqrt{s}=4$ TeV) is shown in Fig. 3. 
 
At LHC ($pp$, $\sqrt{s}=14$ TeV), $Q\bar Q$ pairs can be copiously 
produced.  The 
total cross sections, the gluon-gluon contributions and quark-antiquark 
contributions are shown in Fig. 4 for $M= 200$ GeV to 1000 GeV.  As expected, 
at LHC, the cross sections are dominated the gluon gluon contributions.  For 
$M=200$ GeV, the total cross section $\si$ is $5.6\times 10^4$ pb, while for 
$M= 1$ TeV, $\si=0.90$ pb.  With the projected luminosity \cite{rev} of 
$1.0\times$ 
$10^{34}cm^{-2}\sec^{-1}$, we shall have about 20 billion $Q\bar Q$ 
events for $M=200$ GeV and about 300,000 for $M=1$ TeV. 
 
\section{Production in Photon-Photon Collisions} 
 
\indent\indent  In photon-photon collisons, only the $t$ and $u$-channel of 
fig. 
1 contribute for the $Q\bar Q$ pair production.  For the total cross section, 
we obtain 
\beqra 
\si(\gamma\gamma\rta Q\bar Q) &=& \frac{\pi\al^2}{1215\,\hat s}\;\Bigg\{ 60 
\ln\,\frac{1+\beta}{1-\beta}\;\bigg[15 y^2-8y-22 \nonumber \\ 
&&+ 648\,\frac{1}{y}-1,296\, \frac{1}{y^2}\bigg] + \beta \bigg[3\,y^4+136\, 
y^3-2,772\, y^2 \nonumber \\ 
&& + 6,080\, y-9,720-38,880\,\frac{1}{y}\bigg]\Bigg\} 
\eeqra 
where $\al$ is the fine structure constant to be evaluated at $q^2=M^2$. 
 
The results for the total cross sections for $M=200$ GeV to 1 TeV are given 
in fig. 5 for $\sqrt{s}$ from 500 GeV to 2.5 TeV. 
 
\section*{Concluding Discussion} 
 
\indent\indent  We have calculated the production of exotic spin $\frac{3}{2}$ 
color triplet quarks at high energy hadronic and photon-photon colliders.  At 
LHC, the cross sections are very large and such a particle of mass up to 1 TeV 
will be copicously produced.  At the upgraded Tevatron,   the cross 
sections are somewhat smaller  but still could be observable if 
the mass is 200 GeV or smaller.  We have also calculated the production 
cross sections for photon-photon collisions at various center of mass energies. 
 
  In writing  the interaction  (9), we have assumed that the spin 
$\frac{3}{2}$ quarks are point like.  As a result, at very high energy, our 
cross 
sections grow like $s^3$.  This will violate tree unitarity for large 
enough $s$ 
and so the higher order corrections to our tree diagrams will be important at 
large $s$.  We have taken the point of view that  (9) represents an 
effective 
interaction, and at very high energy, the cross section will be damped due to 
some form factors.   As a result our cross sections are probably 
over estimated.  Also,  in order to detect the signals 
for the spin $\frac{3}{2}$ quark production, we need to know how the particles 
decay.  We have not discussed signals here, since it is addressed in Ref. 2. 
 
\section*{Acknowledgements} 
 
\indent\indent This work was supported in part by U.S. Department of Energy 
Grants\break DE-FG13-93ER40757 and DE-FG02-94ER40852.

\pagebreak

\vspace{24pt} 
 
\noindent 
\Large{\bf Figure~ Captions} 
 
\normalsize 
 
\vspace{14pt} 
 
\begin{enumerate} 
\normalsize 
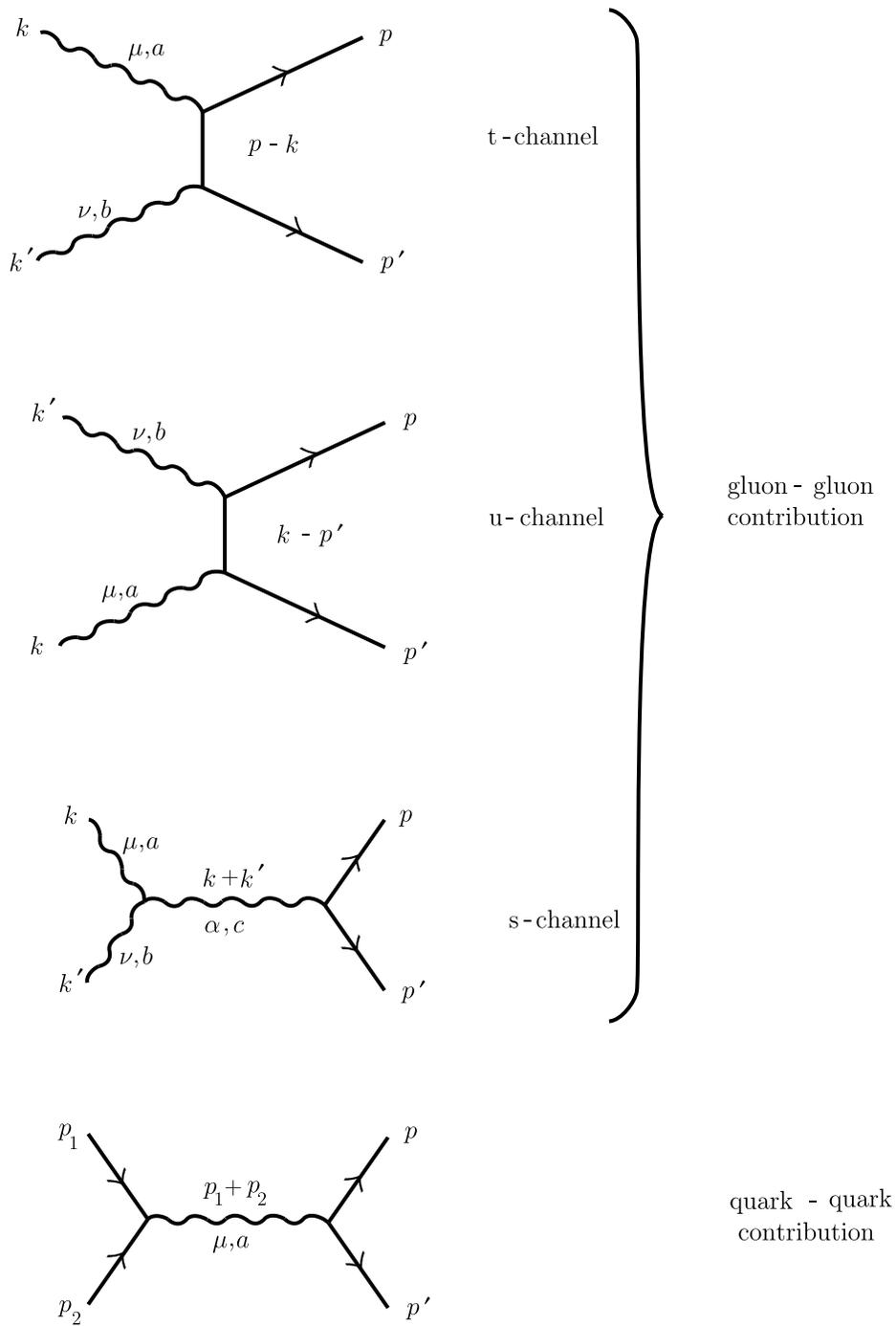
\item[{[Fig 1]}]  The tree diagrams for the sub-processes that contribute 
to the 
production of heavy spin $\frac{3}{2}$ quark pairs in $pp$ and $\bar p p$ 
collisions. 
 
\item[{[Fig 2]}] The cross section as a function of mass for spin 
$\frac{3}{2}$ 
quark pairs at the Tevatron $(\sqrt{s}= 1.8$ TeV) showing the quark-quark 
contribution (top line) and the gluon-gluon contribution (bottom line). 
The top 
line is also the total cross section. 
 
\item[{[Fig 3]}] The spin $\frac{3}{2}$ quark pair cross section as a 
function of 
mass at TeV 2000 $(\sqrt{s}=4$ TeV).  The middle line is the quark-anti-quark 
contribution.  The bottom, the gluon-gluon contribution.  The top line is 
the total. 
 
\item[{[Fig 4]}] The spin $\frac{3}{2}$ quark pair cross section as a 
function of 
mass at LHC $(\sqrt{s}=14$ TeV).  The middle line is the gluon-gluon 
contribution. 
The bottom line is the quark anti-quark contribution.  The top line is the 
total. 
 
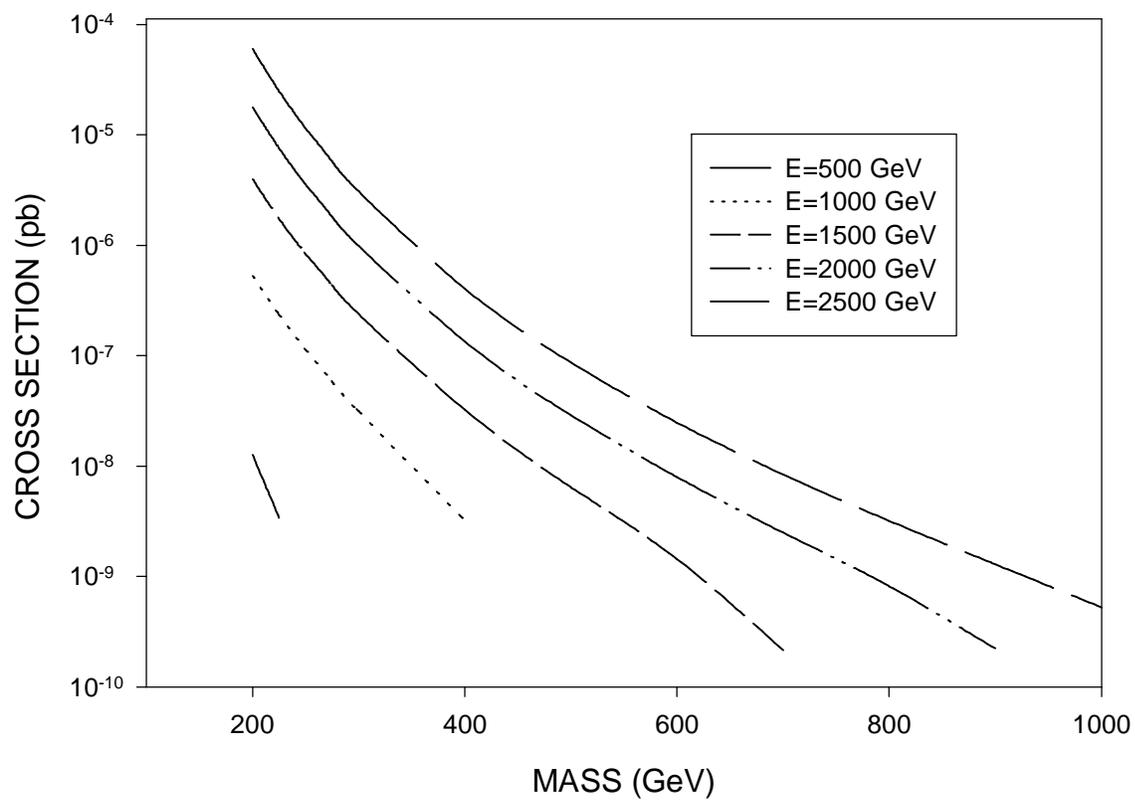
\item[{[Fig 5]}] The spin $\frac{3}{2}$ quark pair cross section for 
photon-photon 
reactions for different center of mass energies as a function of the spin 
$\frac{3}{2}$ quark mass. 
 
\end{enumerate} 
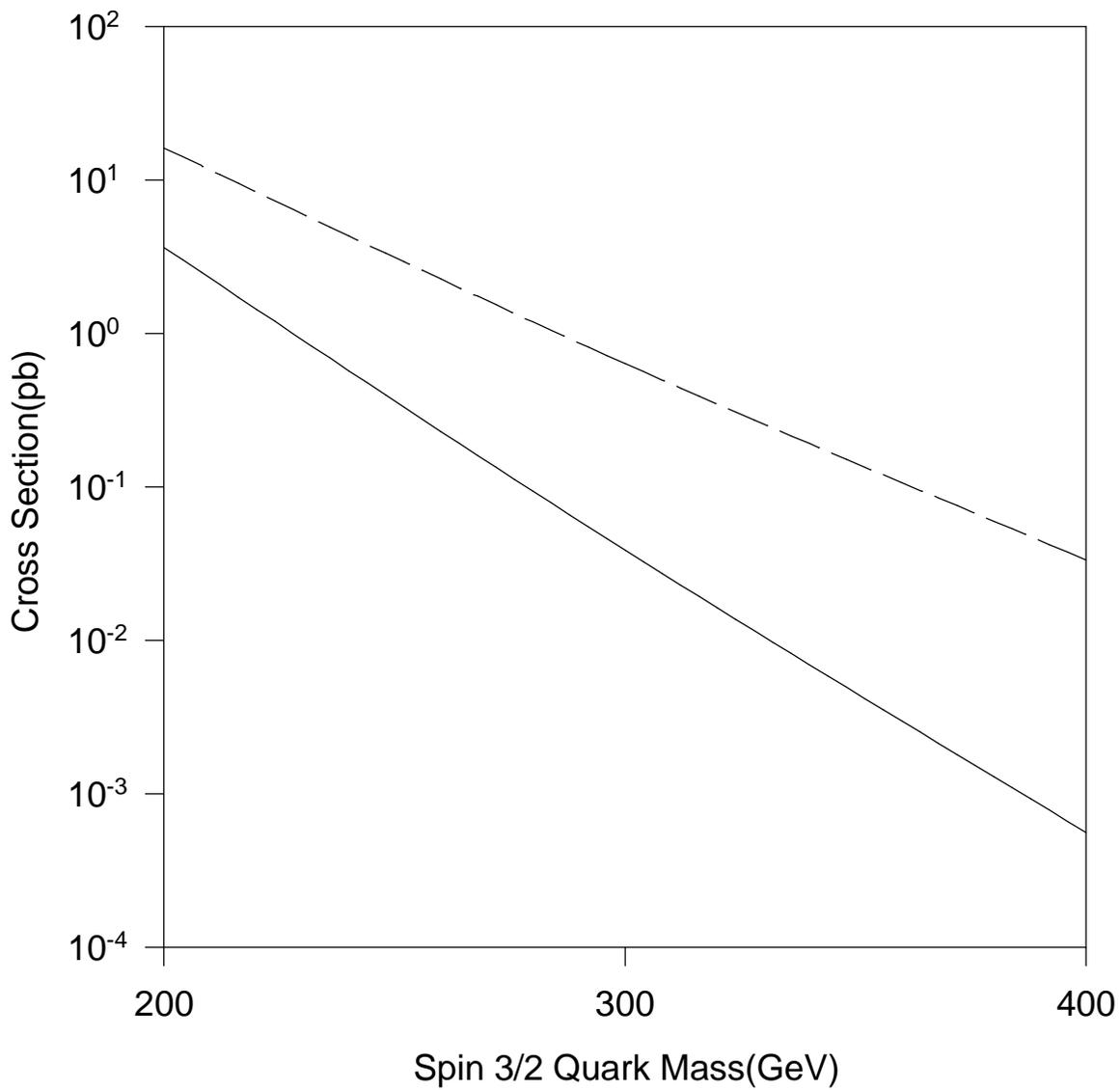
\end{document}